\documentclass[letterpaper,11pt]{article}
\usepackage{fullpage}

\newtheorem{theorem}{Theorem}
\newcommand{\occ}{\ensuremath{\mathrm{occ}}}

\begin{document}

\title{Worst-case optimal adaptive alphabetic prefix-free coding}
\author{Travis Gagie\thanks{Funded by NSERC Discovery Grant RGPIN-07185-2020.  We presented a related result~\cite{Gag22} about the non-alphabetic case at the 30th European Symposium on Algorithms (ESA '22).  This manuscript was checked by AI.}\\
0000-0003-3689-327X\\[1ex]
Faculty of Computer Science\\
Dalhousie University\\
Halifax, Canada}
\maketitle

\begin{abstract}
\noindent
We give the first algorithm for adaptive alphabetic prefix-free coding that is worst-case optimal in terms of time and compression when $\sigma \in o \left( \frac{n^{1 / 2}}{\log n} \right)$, where $\sigma$ is the size of the alphabet and $n$ is the length of the input.

\bigskip

\end{abstract}

\section{Introduction}
\label{sec:introduction}

Suppose we want to encode and transmit over a noiseless binary channel a string of length $n$ over the alphabet $\{1, \ldots, \sigma\}$ such that a receiver can decode each character as soon as they receive its last bit.  Moreover, we want to make a single pass over the string, encoding and sending each character before reading the next one.  Finally, we want the lexicographic order of the possible strings' encodings to be the same as that of the possible strings themselves.  How long must encoding and decoding take in the worst case and how many bits must we send, in terms of $n$, $\sigma$ and the entropy of the characters' distribution?

This is the alphabetic version of the classic problem of adaptive prefix-free coding.  In Section~\ref{sec:bounds} we review previous worst-case bounds for both the non-alphabetic and alphabetic versions.  In Section~\ref{sec:algorithm} we give a simple algorithm for adaptive alphabetic prefix-free coding.  In Section~\ref{sec:analysis} we show this algorithm is worst-case optimal in terms of time and compression when $\sigma \in o \left( \frac{n^{1 / 2}}{\log n} \right)$.  The best previously known algorithm for this problem uses $O (\log \log n)$ time per character to decode, whereas ours uses constant time per character.  We assume throughout that we are working on a word-RAM with $\Omega (\log n)$-bit words.

In a recent survey of alphabetic coding, Bruno, De Prisco and Vaccaro~\cite{BDV25} observed that
\begin{quotation}
\noindent
``At the best of our knowledge, there are no algorithms that efficiently update the structure of optimal alphabetic codes, as the symbol probabilities change.  It would be interesting to design such algorithms, in the same spirit of what has been done by Knuth for classical Huffman codes.''
\end{quotation}
Our algorithm and its bounds do not address their observation directly because it is based on Gilbert and Moore's~\cite{GM59} construction of alphabetic prefix-free codes, which is a modification of Shannon's~\cite{Sha48} construction and generally sub-optimal in the non-adaptive case; moreover, we update our code only intermittently.  Nevertheless, we hope it provides insight.

Although adaptive Shannon coding beats adaptive Huffman coding in the worst case, the opposite is usually true in practice --- as in the non-adaptive case --- since real inputs are better modelled as coming from a probabilistic source than from an adversary.  Therefore, although our algorithm is simple, we consider it of more theoretical than practical interest and have not implemented it.

\section{Previous bounds}
\label{sec:bounds}

In 1973 Faller~\cite{Fal73} showed how we can encode each character in a string using a Huffman code~\cite{Huf52} for the distribution of characters we have already seen (modified to assign codewords also to unseen characters), such that encoding and decoding each character and updating the code afterwards takes time proportional to the length of that character's codeword.  This allows us to encode using a single pass, instead of using a first pass to find the characters' frequencies and a second pass to encode.  Gallager~\cite{Gal78} and Knuth~\cite{Knu85} further developed Faller's algorithm in 1978 and 1985, respectively, and thus it is usually known as the FGK algorithm.

In 1987 Vitter~\cite{Vit87} gave a more sophisticated algorithm for one-pass Huffman coding --- better known as adaptive (or dynamic) Huffman coding --- and showed it uses less than 1 more bit per character than two-pass non-adaptive Huffman coding.  This means that for a string $S [1..n]$ over the alphabet $\{1, \ldots, \sigma\}$, Vitter's algorithm uses at most about $H + 1 + \delta$ bits per character, where
\[H = \sum_{i = 1}^\sigma \frac{\occ (i, S)}{n} \cdot \lg \frac{n}{\occ (i, S)}
\leq \lg \sigma\]
is the entropy of the distribution $\frac{\occ (1, S)}{n}, \ldots, \frac{\occ (\sigma, S)}{n}$ of characters in the string, $\delta \in [0, 1)$ is the redundancy of a Huffman code for that distribution and $\occ (i, S)$ denotes the frequency of $i$ in $S$.  (Throughout this paper, we use $\lg$ for the binary logarithm and $\log$ inside asymptotic notation when the base does not matter.)  Vitter's algorithm also encodes and decodes each character and updates the code afterwards in time proportional to the length of that character's codeword, meaning it uses $O (H + 1)$ time per character to encode and decode.

Vitter attributed to Chazelle an observation implying the FGK algorithm uses at most about twice as many bits as non-adaptive Huffman coding (the coefficient 2 can be reduced to about $1.44$ using a result by Katona and Nemetz~\cite{KN76}), so the FGK algorithm also uses $O (H + 1)$ time per character to encode and decode.  In 1999 Milidi\'u, Laber and Pessoa~\cite{MLP99} showed that FGK uses less than 2 more bits per character than non-adaptive Huffman coding, so at most about $H + 2 + \delta$ bits per character.

In 2004 Gagie~\cite{Gag04} showed how we can modify the FGK algorithm to perform adaptive Shannon coding instead of adaptive Huffman coding, and use at most about $H + 1$ bits per character.  A Shannon code~\cite{Sha48} is a prefix-free code that assigns any character with probability $p$ a codeword of length at most $\left\lceil \lg \frac{1}{p} \right\rceil$.  Like FGK, his algorithm uses $O (H + 1)$ time per character to encode and decode.  In 2009 Karpinski and Nekrich~\cite{KN09} gave a more sophisticated algorithm for adaptive Shannon coding, using canonical coding~\cite{Lee76}, that still uses at most about $H + 1$ bits per character but only constant time per character for encoding and $O (\log (H + 2))$ time per character for decoding.  Shortly thereafter, Gagie and Nekrich~\cite{GN09} improved Karpinski and Nekrich's algorithm so that decoding also takes constant time per character, although they required that $\sigma \in o \left( \frac{n}{\log^{5 / 2} n} \right)$.

Gagie and Nekrich proved we must use at least about $\lg \sigma + 1 \geq H + 1$ bits in the worst case, even when $\sigma \in \omega (1)$.  To see why, suppose $\sigma = 2^{\lceil \lg f (n) \rceil} + 1$ for some function $f (n) \in \omega (1)$, so $\sigma \in \omega (1)$ and any prefix-free code for the alphabet assigns some character a codeword of length at least $\lceil \lg f (n) \rceil + 1 = \lg \sigma + 1 - o (1)$.  For each $j$, the adversary chooses $S [j]$ to be a character with codeword length at least $\lg \sigma + 1 - o (1) \geq H + 1 - o (1)$ in the code we use to encode $S [j]$.  Therefore, their algorithm is worst-case optimal in terms of encoding time, decoding time and compression.

Gagie showed how to modify his algorithm for adaptive alphabetic prefix-free coding, by using Gilbert and Moore's~\cite{GM59} modification of Shannon's construction.  Gilbert and Moore's construction produces an alphabetic prefix-free code --- that is, one in which the lexicographic order of the codewords is the same as that of the characters --- that assigns any character character with probability $p$ a codeword of length at most $\left\lceil \lg \frac{1}{p} \right\rceil + 1$.  Gagie's modified algorithm uses at most about $H + 2$ bits per character and $O (H + 1)$ time per character to encode and decode.  Gagie and Nekrich sped up Gagie's algorithm to use only constant time per character for encoding and $O (\log \log n)$ time per character for decoding, again requiring that $\sigma \in o \left( \frac{n}{\log^{5 / 2} n} \right)$.

If $\sigma = 2^{\lceil \lg f (n) \rceil} + 1$ for some function $f (n) \in \omega (1)$, then any alphabetic prefix-free code for the alphabet assigns two lexicographically consecutive characters codewords of length at least $\lceil \lg f (n) \rceil + 1 = \lg \sigma + 1 - o (1)$.  Therefore, for each $j$, the adversary can choose $S [j]$ to be an even character with codeword length at least $\lg \sigma + 1 - o (1)$ in the code we use to encode $S [j]$.  Since the adversary always chooses even characters, $H \leq \lg \sigma - 1$, so Gagie and Nekrich's algorithm for adaptive alphabetic prefix-free coding is worst-case optimal in terms of encoding time and compression, just not decoding time.

The per-character bounds for the algorithms discussed here (and another by Golin et al.~\cite{GILMN18}) and our new algorithm are shown in Table~\ref{tab:bounds}, assuming $\sigma \in o \left( \frac{n^{1 / 2}}{\log n} \right)$ and ignoring lower-order terms.

\begin{table}
\begin{center}
\caption{The per-character bounds for the algorithms discussed, assuming $\sigma \in o \left( \frac{n^{1 / 2}}{\log n} \right)$ and ignoring lower-order terms.  Algorithms for adaptive prefix-free coding are in the top section and the ones for adaptive alphabetic prefix-free coding are in the bottom section.  We could not determine the decoding time of Golin et al.'s algorithm.}
\label{tab:bounds}

\bigskip

\begin{tabular}{r|ccc}
& encoding & encoding & decoding\\
authors & length & time & time\\
\hline\\[-1.5ex]
FGK~\cite{Fal73,Gal78,Knu85} & $H + 2 + \delta$ & $O (H + 1)$ & $O (H + 1)$\\
Vitter~\cite{Vit87} & $H + 1 + \delta$ & $O (H + 1)$ & $O (H + 1)$\\
Gagie~\cite{Gag04} & $H + 1$ & $O (H + 1)$ & $O (H + 1)$\\
KN~\cite{KN09} & $H + 1$ & $O (1)$ & $O (\log (H + 2))$\\
GN~\cite{GN09} & $H + 1$ & $O (1)$ & $O (1)$\\[1ex]
\hline\\[-1.5ex]
Gagie~\cite{Gag04} & $H + 2$ & $O (H + 1)$ & $O (H + 1)$\\
GN~\cite{GN09} & $H + 2$ & $O (1)$ & $O (\log \log n)$\\
GILMN~\cite{GILMN18} & $H + O (1)$ & $O (1)$ & ?\\
Theorem~\ref{thm:new} & $H + 2$ & $O (1)$ & $O (1)$
\end{tabular}
\end{center}
\end{table}

\section{Algorithm}
\label{sec:algorithm}

We process $S [1..n]$ in blocks of non-decreasing size and encode each block character by character with a code based on all the previous blocks.  The first block is $S [1..\sigma]$ and, for $j \leq \sigma$, we encode $S [j]$ as its $\lceil \lg \sigma \rceil$-bit binary representation.  If a block ends at $S [b]$ then the next block is $S [b + 1..\min (b + \lceil \sigma \lg b \rceil, n)]$.  To encode that next block, we use Gilbert and Moore's~\cite{GM59} construction to build an alphabetic prefix-free code for the probability distribution
\[\frac{\lg b - 1}{\lg b} \cdot \frac{\occ (1, S [1..b])}{b} +
	\frac{1}{\lg b} \cdot \frac{1}{\sigma}, \ldots,
	\frac{\lg b - 1}{\lg b} \cdot \frac{\occ (\sigma, S [1..b])}{b} +
	\frac{1}{\lg b} \cdot \frac{1}{\sigma}\,,\]
which is a weighted average of the distribution of characters we have seen so far and the uniform distribution.

Since Gilbert and Moore's construction assigns any character with probability $p$ a codeword of length at most $\left\lceil \lg \frac{1}{p} \right\rceil + 1$ and every character is assigned probability at least $\frac{1}{\lg b} \cdot \frac{1}{\sigma}$ in the distribution above, the maximum codeword length is at most $\lceil \lg (\sigma \lg b) \rceil + 1$.  Therefore, we can build one lookup table with $\sigma$ entries for encoding and another with $O \left( 2^{\lceil \lg (\sigma \lg b) \rceil + 1} \right) = O (\sigma \log b)$ entries for decoding.  The encoding table tells us the codeword for any character and the decoding table tells us, for any string of $\lceil \lg (\sigma \lg b) \rceil + 1$ bits, which character's codeword is a prefix of that string and the length of that codeword; each entry in the tables takes a constant number of machine words.

We encode the next block character by character with this code, using the lookup tables to encode and decode each character in constant time.  Notice we need not know $n$ in advance.

\section{Analysis}
\label{sec:analysis}

We encode and decode the first block $S [1..\sigma]$ in constant time per character.  If a block ends at $S [b]$ then building the code and lookup tables for the next block takes $O (\sigma \log b)$ time.  If $b + \lceil \sigma \lg b \rceil \leq n$ then we charge this to the $\lceil \sigma \lg b \rceil$ characters in the next block; otherwise, we consider it a one-time $O (\sigma \log n)$ overhead.  It follows that we use $O (n + \sigma \log n)$ time overall for both encoding and decoding.

We encode each character using at most $\lceil \lg (\sigma \lg n) \rceil + 1$ bits, so we encode the first $\lceil \sigma \lg n \rceil$ occurrences of each distinct character using $O ((\sigma \log \max (n, \sigma))^2)$ total bits.  Suppose a block ends at $S [b]$ and consider a character $S [j]$ in the next block that is not one of the first $\lceil \sigma \lg n \rceil$ occurrences of that distinct character, meaning $\occ (S [j], S [1..j]) > \lceil \sigma \lg n \rceil$.  Since
\begin{eqnarray}
\lefteqn{\occ (S [j], S [1..b])} \\
& \geq & \occ (S [j], S [1..j]) - (j - b) \\
& \geq & \occ (S [j], S [1..j]) - \lceil \sigma \lg n \rceil
\end{eqnarray}
and $(1 + 1 / x)^x < e$ for $x > 0$, we encode $S [j]$ using at most
\begin{eqnarray}
\lefteqn{\left\lceil \lg \left( \frac{\lg b}{\lg b - 1} \cdot \frac{b}{\occ (S [j], S [1..b])} \right) \right\rceil + 1} \\
& < & \lg b - \lg \occ (S [j], S [1..b]) +
  \frac{1}{\lg b - 1} \lg \left( 1 + \frac{1}{\lg b - 1} \right)^{\lg b - 1}
  + 2 \\
& < & \lg j - \lg \left( \rule{0ex}{2ex} \occ (S [j], S [1..j]) - \lceil \sigma \lg n \rceil \right) + \frac{\lg e}{\lg b - 1} + 2
\label{line:first_split}
\end{eqnarray}
bits.  Since $\sigma \leq b < j \leq b + \lceil \sigma \lg b \rceil$, we have $j < b + b \lg j + 1$ and so $b > \frac{j - 1}{\lg j} - 1$.  Therefore, the sum $\frac{\lg e}{\lg b - 1} + 2$ of the third and fourth terms in line~(\ref{line:first_split}) goes to 2 as $j$ increases, and the sum over $j$ of those terms is $n (2 + o (1))$.

Consider the sum of the first two terms in line~(\ref{line:first_split}) over all characters $S [j]$ with $\occ (S [j], S [1..j]) > \lceil \sigma \lg n \rceil$.  Since
\begin{eqnarray}
\lefteqn{\lg ((\occ (i, S) - \lceil \sigma \lg n \rceil)!)} \\
& = & \lg (\occ (i, S)!) - \sum_{k = 0}^{\lceil \sigma \lg n \rceil - 1} \lg (\occ (i, S) - k) \\
& \geq & \lg (\occ (i, S)!) - O (\sigma \lg^2 n)\,,
\end{eqnarray}
we have
\begin{eqnarray}
\lefteqn{\sum \left\{ \lg j - \lg \left( \rule{0ex}{2ex} \occ (S [j], S [1..j]) - \lceil \sigma \lg n \rceil \right)\ :\ \occ (S [j], S [1..j]) > \lceil \sigma \lg n \rceil \right\}} \\
& < & \lg (n!) - \sum \left\{ \lg \left( \rule{0ex}{3ex} \left( \rule{0ex}{2ex} \occ (i, S) - \lceil \sigma \lg n \rceil \right) ! \right)\ :\ \occ (i, S) > \lceil \sigma \lg n \rceil \right\} \\
& \leq & \lg (n!) - \sum_{i = 1}^\sigma \lg (\occ (i, S)!) + O ((\sigma \log n)^2)\,.
\label{line:second_split}
\end{eqnarray}
By Stirling's Approximation, the binary logarithm of the number of distinct arrangements of the characters in $S$
\begin{eqnarray}
\lg (n!) - \sum_{i = 1}^\sigma \lg (\occ (i, S)!)
\leq n H + O (\sigma \log n)\,.
\label{line:Stirlings}
\end{eqnarray}

Adding together the $O ((\sigma \log \max (n , \sigma))^2)$ total bits we use for the $\lceil \sigma \lg n \rceil$ first occurrences of each distinct character, the $n (2 + o (1))$ from summing over $j$ the third and fourth terms in line~(\ref{line:first_split}), the $O ((\sigma \log n)^2)$ term in line~(\ref{line:second_split}), and the $n H + O (\sigma \log n)$ bound from line~(\ref{line:Stirlings}), we get our result:

\begin{theorem}
\label{thm:new}
Our algorithm for adaptive alphabetic prefix-free coding encodes $S [1..n]$ using
\[n (H + 2 + o (1)) + O ((\sigma \log \max (n, \sigma))^2)\]
bits, which is at most about $H + 2$ bits per character when $\sigma \in o \left( \frac{n^{1 / 2}}{\log n} \right)$, and constant time per character for encoding and decoding when $\sigma \in O (n / \log n)$.
\end{theorem}

\noindent
If we use Shannon's construction instead of Gilbert and Moore's then we get an algorithm~\cite{Gag22} for adaptive prefix-free coding, without the alphabetic constraint, that is simpler than Gagie and Nekrich's~\cite{GN09} and worst-case optimal when $\sigma \in o \left( \frac{n^{1 / 2}}{\log n} \right)$.

% thank Jeff, Ugo, Nando...?

\end{document}